\documentclass[english,aps,prb,twocolumn, superscriptaddress,showpacs, amsmath]{revtex4-1}

\usepackage[T1]{fontenc}
\usepackage[latin9]{inputenc}
\usepackage{color}
\usepackage{units}
\usepackage{amssymb}
\usepackage{graphicx}
\usepackage{esint}
\usepackage{bm}
\usepackage{natbib}
\usepackage{amsmath}
\usepackage[normalem]{ulem}

\setcitestyle{journalcolor= blue}

\usepackage{pdfpages}
\usepackage{babel}

\makeatletter
\AtBeginDocument{\let\LS@rot\@undefined}
\makeatother

\begin{document}
\title{Enhanced shot noise at bilayer graphene - superconductor junction}

\author{Manas Ranjan Sahu}
\thanks{equally contributed}
\affiliation{Department of Physics, Indian Institute of Science, Bangalore 560012, India}
\author{Arup Kumar Paul}
\thanks{equally contributed}
\affiliation{Department of Physics, Indian Institute of Science, Bangalore 560012, India}
\author{Abhiram Soori}
\affiliation{Department of Physics, Indian Institute of Science, Bangalore 560012, India}
\affiliation{School of Physics, University of Hyderabad, Gachibowli, Hyderabad 500046, India}
\author{K. Watanabe}
\affiliation{National Institute for Materials Science, 1-1 Namiki, Tsukuba 305-0044, Japan}
\author{T. Taniguchi}
\affiliation{National Institute for Materials Science, 1-1 Namiki, Tsukuba 305-0044, Japan}
\author{Subroto Mukerjee}
\affiliation{Department of Physics, Indian Institute of Science, Bangalore 560012, India}
\author{Anindya Das}
\email{anindya@iisc.ac.in}
\affiliation{Department of Physics, Indian Institute of Science, Bangalore 560012, India}

\begin{abstract}
{Transport properties of graphene - superconductor junction has been studied extensively to understand the interplay of the relativistic Dirac quasiparticles and superconductivity. Though shot noise measurements in graphene has been performed to realize many theoretical predictions, both at zero magnetic field as well as quantum Hall (QH) regime, its junction with superconductor remain unexplored. Here, we have carried out the shot noise measurements in an edge contacted bilayer graphene - Niobium superconductor junction at zero magnetic field as well as QH regime. At the Dirac point we have observed a Fano factor $\sim 1/3$  above the superconducting gap ($\Delta$) and a transition to an enhanced Fano factor $\sim 0.5$ below the superconducting gap. By changing the carrier density we have found a continuous reduction of Fano factor for both types of carriers, however the enhancement of Fano factor within the superconducting gap by a factor of $\sim 1.5$ is always preserved. The enhancement of shot noise is also observed in the QH regime, where the current is carried by the edge state, below the critical magnetic field and within the superconducting gap. These observations clearly demonstrate the enhanced charge transport at the graphene-superconductor interface.
}
\end{abstract}

\maketitle

\section{INTRODUCTION}

Engineering a topological superconductor, which can host exotic non-abelian Majorana quasi-particles\cite{fu2008superconducting}, has been one of the emerging research areas in mesoscopic condensed matter physics\cite{lutchyn2010majorana,oreg2010helical,mourik2012signatures,das2012zero,mong2014universal,clarke2014exotic}. Realizing superconducting correlations in a quantum Hall edge has been proposed as a novel route for creating even more exotic topological entities, such as parafermion\cite{orth2015non} or Fibbonacci particles\cite{mong2014universal}. The conventional conductance measurement is the well developed technique to probe the superconducting correlations. However, in many cases various important information stay hidden in this technique due to its averaging nature. In such cases shot noise measurement has been proven to be an unique tool for studying physics of different kind of interactions\cite{khlus1987current,de1996semiclassical,blanter2000shot,kaviraj2011noise,hata2018enhanced,oberholzer2002crossover,choi2005shot,ota2017negative,beenakker2003quantum,das2012high,steinbach1996observation,lefloch2003doubled}, such as the effective charge ($e^*$) of the quasiparticle. Observing enhanced charge transport ($e^* > e$) at QH - superconductor junction would be a first step in realizing superconducting correlations in the QH edge.

In the last two decades shot noise technique has been extensively used either to understand or to produce many interesting physics, such as one third Fano factor in disordered metal\cite{beenakker1992suppression,nagaev1992shot,steinbach1996observation,henny19991}, fractional charge in Fractional QH regime\cite{saminadayar1997observation,reznikov1999observation,hashisaka2015shot}, charge doubling in normal metal-superconductor (NS) junction\cite{de1994doubled,jehl2000detection,nagaev2001semiclassical}, discrete quantization of charge in multiple Andreev reflection (AR)\cite{ronen2016charge,hoss2000multiple}. In graphene one third Fano factor has been obseved at the Dirac point \cite{tworzydlo2006sub,dicarlo2008shot,danneau2008shot}. Further shot noise measurement is performed in graphene PN junctions at zero magnetic field as well as in the QH regime\cite{kumada2015shot,matsuo2015edge,abanin2007quantized}. However, the shot noise at graphene - superconductor junction is not studied till date to understand the effect of Andreev processes in these hybrids both at zero magnetic field and QH regime.

\begin{figure*}
\begin{center}
\centerline{\includegraphics[width=1\textwidth]{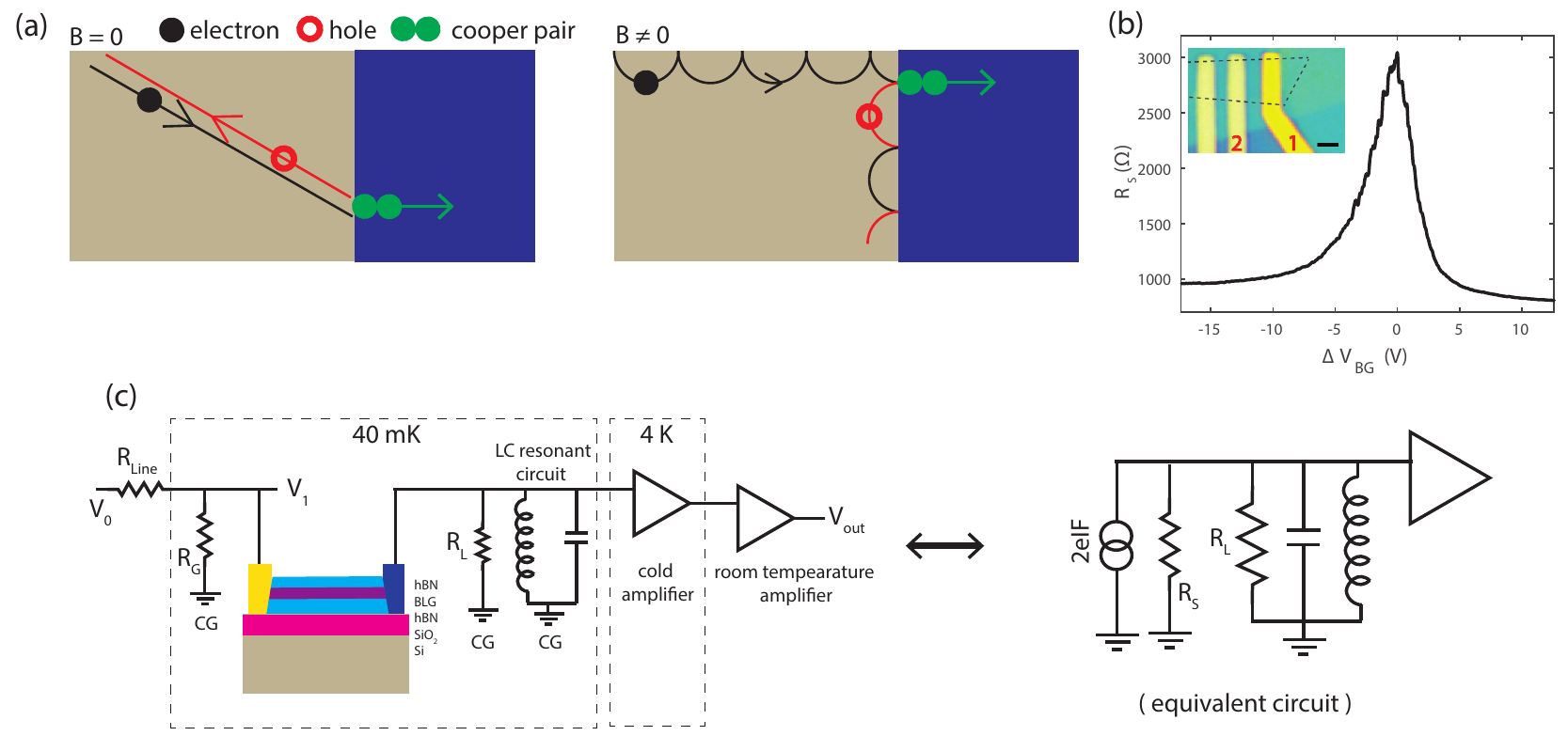}}
\caption{\textbf{Measurement setup.} (\textbf{a}) The schematic of AR at graphene-superconductor junction at zero magnetic field (left panel) and in QH regime (right panel). (\textbf{b}) Resistance of the BLG device plotted as a function of $\Delta V_{BG}$, where the $\Delta V_{BG}$ is the back gate voltage with respect to the Dirac point. The inset shows the optical image of the hBN encapsulated BLG device, where all the measurements were performed between contact-1 (Au) and contact-2 (Nb). Scale bar shown is of 1$\mu$m. (\textbf{c}) The schematic of the device and the shot noise measurement set-up (left panel) with an equivalent circuit (right panel). The resistances used are line resistance ($R_{Line}$) = 65$\Omega$, resistance to cold ground ($R_G$) = 5 $\Omega$, load resistance ($R_L$) = 20 k$\Omega$ and CG is the cold ground. The values of inductance and capacitance that forms the resonant circuit are 365 $\mu$H and 138 $pF$, respectively.}
\label{Figure1}
\end{center}
\end{figure*}

In this work we have carried out the shot noise measurement in a bilayer graphene (BLG) - Niobium (Nb) superconductor junction. We have observed the Fano factor $\sim 1/3$ at the Dirac point for normal quasi-particle transport above the superconducting gap ($|eV_{SD}| > \Delta$). Whereas within the superconducting gap ($|eV_{SD}| < \Delta$), when the transport occurs via AR, an enhanced Fano factor $\sim 0.5$ is observed. The schematic of the Andreev process\cite{sahu2016andreev,soori2018enhanced,shalom2016quantum,lee2018proximity} is shown in Fig.~1a-left panel. By increasing the carrier density we found a continuous reduction of Fano factor for both normal transport as well as Andreev process signifying the quasi-ballistic nature of the device. However, the enhancement of Fano factor by a factor of $\sim 1.5$ due to the Andreev process compared to normal transport was observed at all carrier densities. The AR in the QH regime\cite{rickhaus2012quantum,shalom2016quantum,amet2016supercurrent,lee2017inducing,park2017propagation,sahu2018inter,lee2018proximity} is shown schematically in Fig.~1a-right panel, where the shot noise is being created due to the finite barrier at the interface between the edge state and the superconductor. An enhanced shot noise is also observed at the QH-superconductor junction within the superconducting gap ($|eV_{SD}| < \Delta$) and below the critical magnetic field of the Nb superconductor.  The enhancement of shot noise at the graphene-superconductor junction clearly suggests the enhanced charge transport ($e^* > e$) in our experiment.

\section{DEVICE FABRICATION AND CHARACTERIZATION}
The device consists of an edge contacted Bilayer graphene connected to a Gold (Au) contact at one end and a Niobium superconducting contact at other end. The device was fabricated using standard dry transfer technique\cite{pizzocchero2016hot}. The normal contact was achieved by electron beam lithography followed by dry ion etching and thermal deposition of  Cr/Pd/Au (3/9/68~nm). The superconducting contact was achieved using the same technique and by deposition of Ti/Nb (5/30 nm) using sputtering. The channel length (L) and width (W) of device are $\sim 0.7 \mu m$ and $\sim 2.1 \mu m$, respectively, which makes W/L $\sim$ 3 (Fig. 1(b) inset).

All the measurements were done in a cryofree dilution refrigerator having base temperature of $\sim 40~mK$. The conductance measurement was done using standard lock-in technique at 728 Hz in two probe configuration and we have subtracted the line resistances for all the presented data. The shot noise measurements were performed using LCR resonant circuit at 710 kHz. The gate response of the device is shown in the Fig. 1(b), from which we have extracted the contact resistance $\sim$ 700-800$\Omega$ (Supplementary Fig.~2) of the device. The measured contact resistance is comparatively higher than the typical contact resistance values for graphene devices with Cr/Pd/Au contacts at both sides. The higher contact resistance is possibly due to the Ti/Nb contact. The superconducting gap and the critical magnetic field ($B_C$) of $30~nm$ thick Nb film was characterized separately as shown in Supplementary Fig.~1. The $B_C$ was found to be 4T and from the critical temperature ($T_C \sim$7 K) of the film we found $2\Delta_{Nb} \sim 2~meV$ (=3.528$k_BT_C$). Note that the 30 nm thin Nb film was chosen in order to have higher $B_C$.

\begin{figure*}
\begin{center}
\centerline{\includegraphics[width=1\textwidth]{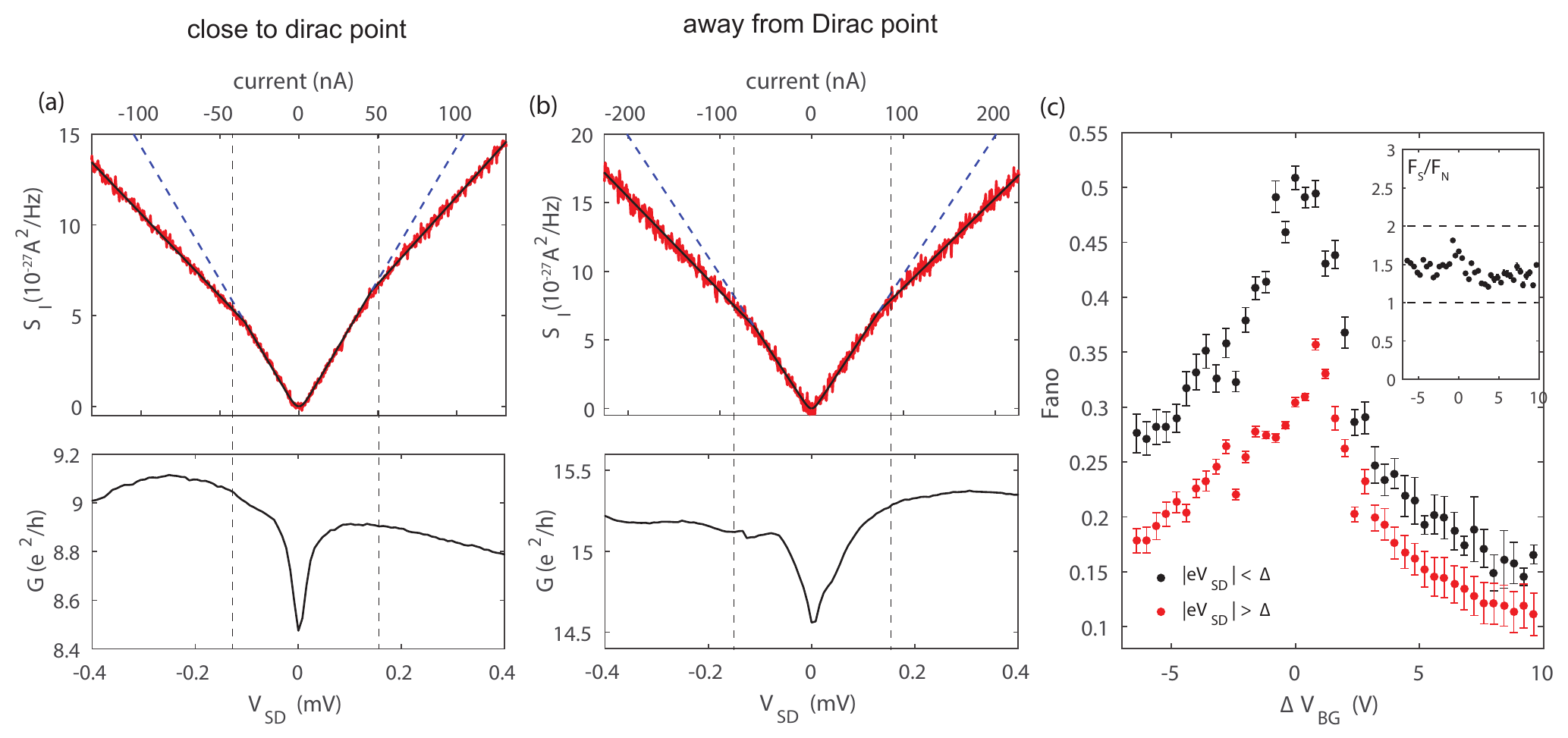}}
\caption{\textbf{Shot noise at B=0.} (\textbf{a}) (top) Shot noise plotted as a function of DC current close to the Dirac point ($\Delta V_{BG} = -0.4~V$). The experimental data is fitted (black solid line) to the equation 2 of the text to get the Fano factor below superconducting gap ($F_S$) and above superconducting gap($F_N$). Two distinct slopes are clearly visible, evaluated Fano factors are $F_S \sim 0.46$ and $F_N \sim 0.28$ respectively. The dashed line shows the enhanced slope inside the superconducting gap. (bottom) Diffrential conductance (G) plotted as a function of $V_{SD}$ at the same gate voltage showing BCS like features, which are characteristics of superconducting gap. (\textbf{b}) (top) Shot noise plotted as a function of DC current away from the Dirac point ($\Delta V_{BG} = -3.4~V$). Evaluated Fano factors are $F_S \sim 0.32$ and $F_N \sim 0.21$ respectively. (bottom) Corresponding differential conductance plot. (\textbf{c}) Fano factor is plotted as a function of gate voltage for both $|eV_{SD}| < \Delta$ and $|eV_{SD}| > \Delta$, a global enhancement of Fano factor is clearly visible within the superconducting gap and the ratio ($F_S/F_N$) is shown in the inset.}
\label{Figure1}
\end{center}
\end{figure*}

\section{Experimental setup}
Fig.~1(c) shows the schematic of the shot noise measurement setup . The voltage fluctuations were measured using LCR resonant circuit tuned to 710 kHz (Supplementary Fig.~3), amplified by a home made preamplifier at 4K plate followed by a room temperature amplifier and finally measured by a spectrum analyzer. The total gain (g $\sim$1300) of the system was measured using a known signal as well as by measuring the thermal noise in the QH regime of graphene. Details of gain calibration is shown in Supplementary Fig.~4.

The measured total noise contains various other noises together with the desired shot noise component ($S_I$, which is the current dependent excess noise), given by\cite{ronen2016charge,Srivastaveaaw5798}
\begin{equation}
 S_V=S_IR^2_{eff} + 4k_BTR_{eff} + S_i^{amp}R^2_{eff} + S_v^{amp}
\end{equation}
where $R_{eff}$ is the parallel resistance of the sample ($R_S$) and the load resistor ($R_L$), $4k_BTR_{eff}$ is the thermal noise, $S_i^{amp}$ and $S_v^{amp}$ are the current and voltage noise of the cold amplifier (CA), respectively. Thus, the shot noise (excess noise) can be extracted from the total noise as, $S_I (I)= \frac{S_V (I)-S_V (I=0)}{R^2_{eff}} $.

\begin{figure*}
\begin{center}
\centerline{\includegraphics[width=0.8\textwidth]{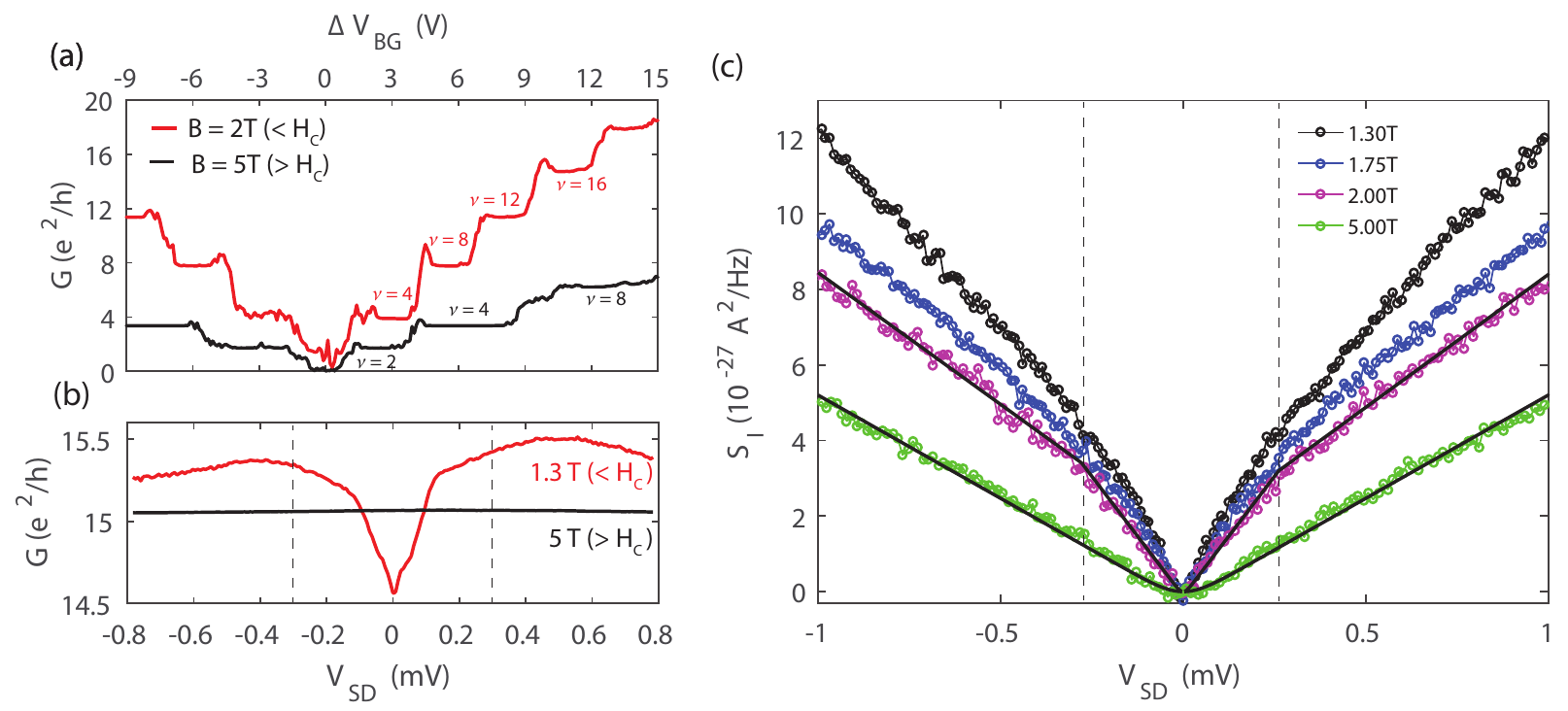}}
\caption{\textbf{Shot noise in QH regime.} (\textbf{a}) Conductance plotted as a function of gate voltage ($\Delta V_{BG}$) at 2~T (below $B_C$) and 5T (abobe $B_C$), respectively showing clear quantized plateaus. Different filling fractions ($\nu$) are mentioned near the corresponding conductance plateaus. The QH data at 5T contains the Nb lead resistance of $\sim 500~\Omega$, which vanishes below $B_C$. The $B_C$ for Nb was observed to be $\sim$ 4~T.  (\textbf{b}) Differential conductance plot at the $\nu = 16$ plateau at B=1.3T (red curve) showing BCS like features, which is not present in the differential conductance plot at the $\nu = 4$ plateau at B=5T (black curve). Note that the black curve is shifted by a conductance value of 11.7 e$^2$/h. (\textbf{c}) Shot noise plotted as a function of bias voltage in a quantum Hall plateau ($\nu = 12$) at several magnetic fields. The colored open circle are the experimental data and the solid lines are the theoretical fittings using equation 2. The presence of higher slopes at $|eV_{SD}|<\Delta$ are apparent for $B < B_C$, where as for $B > B_C$ only a single slope is present.}
\label{Figure1}
\end{center}
\end{figure*}

\begin{figure*}
\begin{center}
\centerline{\includegraphics[width=1\textwidth]{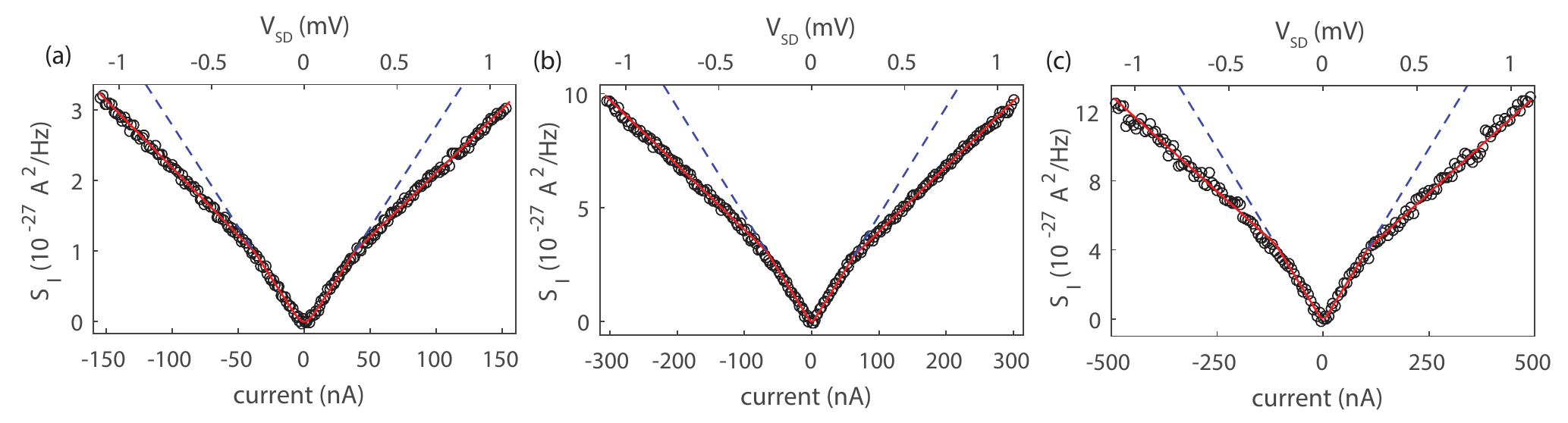}}
\caption{\textbf{Shot noise in QH regime at different B and $\nu$.} (\textbf{a}) Shot noise plotted as a function of DC current at $\nu = 4$ (B=2.5~T), (\textbf{b}) $\nu = 8$ (B=1.75T) and (\textbf{c}) $\nu = 16$ (B=2T). The solid lines are the theoretical fittings using equation 2 to get the Fano factor below superconducting gap ($F_S$) and above superconducting gap ($F_N$). The dashed line shows the enhanced slope inside the superconducting gap.}
\label{Figure1}
\end{center}
\end{figure*}

\section{Shot noise measurement at zero magnetic field}
Shot noise measured at B=0 is shown in Fig.~2 revealing many interesting features. Fig.~2(a) (top) shows the shot noise as a function of applied DC current close to the Dirac point ($\Delta V_{BG} = -0.4~V$) which shows two distinct slopes, one at lower current ($V_{SD} < 0.2~meV$) and other at relatively larger current ($V_{SD} > 0.2~meV$). The outer slope is due to normal quasi-particle transport where as the inner higher slope corresponds to transport by AR. The bias ($V_{SD}$) dependence of differential conductance ($G = 1/R_S$) at the same gate voltage is shown in Fig.~2(a) (bottom) showing BCS like features with a proximity induced gap of 2$\Delta \sim 0.3-0.4~meV$. To extract out the Fano factor, which is the measured shot noise normalized by Poisson noise (2eI), for both normal transport ($F_N$) as well as transport via AR ($F_S$), we have fitted the noise data with the following equations,

\begin{equation}
S_I=\left\{\begin{array}{ll}
2eIF_S \, [coth(\frac{e^*V_{SD}}{2k_BT_e})-\frac{2k_BT_e}{e^*V_{SD}}],& \text{for } |eV_{SD}|<\Delta \\
K + 2eIF_N,              & \text{for } |eV_{SD}|>\Delta
\end{array}\right.\
\end{equation}

where $e$ is the electronic charge, e$^*$=2e,  $k_B$ is the Boltzmann constant, $T_e$ is the electron temperature and $K$ is a constant. Fitting the shot noise data at the Dirac point (Fig. 2a) gives $F_N = 0.3$ and $F_S = 0.51$, respectively. Using the shot noise data the electron temperature ($T_e$) $\sim$ 40 mK was evaluated as shown in Supplementary Fig.~5. The measured $F_N$ value at the Dirac point is close to the theoretically predicted value\cite{tworzydlo2006sub,snyman2007ballistic} as reported in single layer graphene \cite{dicarlo2008shot,danneau2008shot} and attributed to pseudo-diffusive transport by evanescent modes. Ideally, one would expect $F_S = 2F_N$ for an NS interface due to Cooper pair (2e) transport, however, we observe $F_S \sim 1.7F_N$ near the Dirac point. Similar deviation has been reported in topological insulator (TI) - Nb junction\cite{tikhonov2016andreev} and the reduction of Fano factor is attributed to the presence of residual density of states at the interface. Fig.~2(b) shows the shot noise spectrum along with corresponding bias spectrum away from the Dirac point ($\Delta V_{BG} = -3.4~V$), where the fitting gives  $F_N = 0.21$ and $F_S = 0.32$. Fano factors for both $|eV_{SD}|<\Delta$ and $|eV_{SD}|>\Delta$ as a function of $V_{BG}$ are shown in Fig.~2(c), which shows a continuous reduction of Fano factors away from the Dirac point indicating the quasi-ballistic nature of the device\cite{tworzydlo2006sub,snyman2007ballistic}. Inset shows $F_S/F_N$ ratio as a function of $\Delta V_{BG}$, where average enhancement of Fano factor by $\sim$ 1.5 can be clearly seen at all carrier densities.

\section{SHOT NOISE IN THE QH REGIME}

\begin{figure*}
\begin{center}
\centerline{\includegraphics[width=0.7\textwidth]{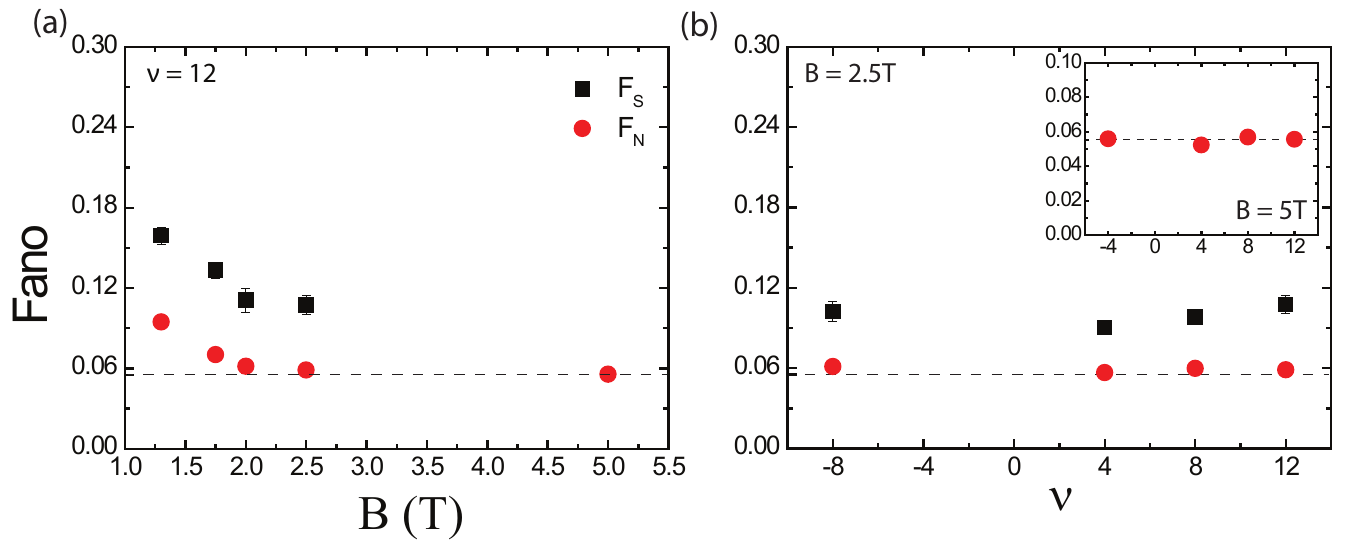}}
\caption{\textbf{Evolution of Fano factor with B and $\nu$.} Black filled squares show the $F_S$ ($|eV_{SD}| < \Delta$) and the red filled circles show the $F_N$ ($|eV_{SD}| > \Delta$). 
(\textbf{a}) Fano factor plotted as a function of B at $\nu = 12$. (\textbf{b}) Fano factor plotted as a function of $\nu$ for $B = 2.5~T$. Here both $F_N$ and $F_S$ are quite independent of $\nu$ indicating presence of cleaner edge states ($E_{LL} > \delta E_{F}$). Inset shows the same for B = 5~T ($B > B_{c}$), where Fano factor is quite independent of $\nu$. Black dashed line in all the figures shows the everage normal state Fano factor at B=5T, $F_N$=0.056.}
\label{Figure1}
\end{center}
\end{figure*}

In order to measure the shot noise in QH regime we applied perpendicular magnetic field and the device starts showing clean quantum Hall plateaus at B as low as $\sim$ 1 T. Fig.~3(a) shows the quantum Hall plateaus at B = 2~T ($B < B_C$) and B = 5~T ($B > B_C$), respectively. Fig.~3(b) shows the bias spectrums measured at the centers of $\nu = 16$ plateau at $B = 1.3~T$ and $\nu = 4$ plateau at $B = 5~T$. The bias spectrum at 1.3T (< B$_C$) shows BCS like features with proximity induced superconducting gap, 2$\Delta \sim 0.4-0.5~meV$, which is not present in the bias spectrum at 5T (> B$_C$). The shot noise data at the center of $\nu = 12$ plateau at several magnetic fields are shown in Fig.~3(c). For a QH edge with fully transparent contacts the shot noise is not expected due to the ballistic nature of the edge state. However, the presence of finite barrier at the interface of quantum Hall edge and the contact will reduce the transmission probability from unity and thus, can create shot noise. The shot noise at 5T with a single slope in Fig.~3(c) is expected for normal quasi-particle transport above the critical field.

The generated shot noise can be quantified by extracting the Fano factors using equation 2. At 5T the $F_N$ $\sim$ 0.056 is extracted from the fitting in Fig.~3(c). Below the critical field, the slope at lower bias voltage ($|eV_{SD}|<\Delta$) is larger compared to the slope at higher bias (($|eV_{SD}|>\Delta$)), similar to the zero magnetic field case. The enhancement of shot noise within the superconducting gap and below the critical magnetic field is the signature of Andreev conversions happening at the QH edge and superconductor interface. Moreover, it can be seen from Fig.~3c that the total noise decreases with increasing B (<$B_C$), both for $|eV_{SD}|<\Delta$ as well as for $|eV_{SD}|>\Delta$. The increment of Fano factors at lower magnetic field could be due to the bulk contribution to the net current, which is expected at lower magnetic field ($B=1.3T$), when the Landau level broadening ($\Gamma$)\cite{sahu2018inter} is comparable to the Landau level gap ($\Delta E_{LL}$). However, it can be seen from Fig.~5a that the Fano factors saturate beyond B=2~T, as the Landau level gap becomes higher with increasing magnetic field. This saturating behavior signifies that beyond B=2~T, the dominant shot noise contribution comes from the edge transport. Similar shot noise data at several other plateaus along with fitted curves using equation 2 is shown in Fig.~4, clearly showing an enhanced shot noise for $|eV_{SD}|<\Delta$ due to Andreev reflections.

Fig.~5(a) shows the $F_S$ and $F_N$ plotted as a function of magnetic field for $\nu=12$ filling factor, where the saturating behavior of both the $F_S$ and $F_N$ is apparent. The $F_S$ and $F_N$ for different filling factors at B=2.5~T are shown in Fig.~5b, where both the Fano factors are pretty constant with an enhancement factor $\sim 2$. The figure 5(b) inset shows the $F_N$ for different filling factors at B=5~T.

Currently, there is no well developed theory for shot noise at the interface of QH edge and a superconductor. Some insights can be gained by considering a ballistic system containing a finite number of modes connected to a superconductor. In such a system, when the superconductor is in normal state, a finite contact resistance can generate shot noise with Fano factor, F$_N$=1-t$_N$, where t$_N$ is the transmittance of the ballistic modes. When the lead becomes superconducting the transmittance and the Fano factor will take the form\cite{blanter2000shot} t$_S$= t$^2_N$/(2-t$_N$)$^2$ and \cite{choi2005shot} F$_S$=8(1-t$_N$)/(2-t$_N$)$^2$, respectively. The ratio of the Fano factors, F$_S$/F$_N$=8/(2-t$_N$)$^2$, will take the minimum value of 2 in the tunneling regime (t$_N$ tends to zero) whereas it will take the maximum value of 8 in the highly transparent regime (t$_N$ tends to one). This suggests that the enhancement of shot noise at the QH-superconductor junction is not universal. Moreover, a finite magnetic field can complicate the problem further due to the presence of inevitable vortices in the superconducting lead. The presence of votices can create pathways for normal quasiparticle transport\cite{miyoshi2005andreev} at the junction, and thus can reduce the ratio of Fano factors from the expected range of values of 2 to 8.

The measured enhancement of Fano factor by $\sim$2 times (Fig. 5) in our highly transparent junction (conductance data in Fig. 3a) is much lesser than the expected value of 8. More theoretical understanding is necessary to quantify the shot noise at the QH-superconductor junction considering the complex nature of dynamics (Fig. 1a - right panel) as well as the presence of vortices. Nevertheless, the enhancement of Fano factor by a factor of $\sim 2$ indicates the enhanced charge transport due to Andreev processes at the QH-superconductor junction, but it does not establish the doubling of charge (e$^*$=2e). We would like to note that there could be noise contribution coming from the normal-QH junction of the device (normal-QH-superconductor), however, that component will not produce a bias dependent slope, and thus will not affect the conclusion of observing enhanced charge transport in our experiment.

\section{CONCLUSION}
In conclusion, we have performed the shot noise measurement in a BLG - superconductor junction for the first time. Below the superconducting gap, when the current is carried via Andreev reflections, enhancement of Fano factor by $\sim 1.5$ times is observed irrespective of carrier densities. More importantly, we have also observed an enhanced shot noise by $\sim 2$ times in the QH regime below the superconducting gap and below the critical magnetic field. The enhancement of shot noise in our experiment clearly signifies the enhanced charge transport at BLG-superconductor junction. However, more theoretical studies are required to understand and quantify the shot noise due to Andreev processes at the quantum Hall - superconductor interface. We believe that our work will pave the way for more investigations utilizing shot noise measurement technique as a tool to study quantum Hall - superconductor hybrids.

\section{ACKNOWLEDGMENTS}
AD thanks DST Nanomission (DSTO2051) for financial support. AS thanks DST Nanomission (DSTO1597) for funding.

\bibliography{references}{}

\onecolumngrid
\newpage
\thispagestyle{empty}
\mbox{}


\section*{Supplementary material}

# SUPPLEMENTARY INFORMATION

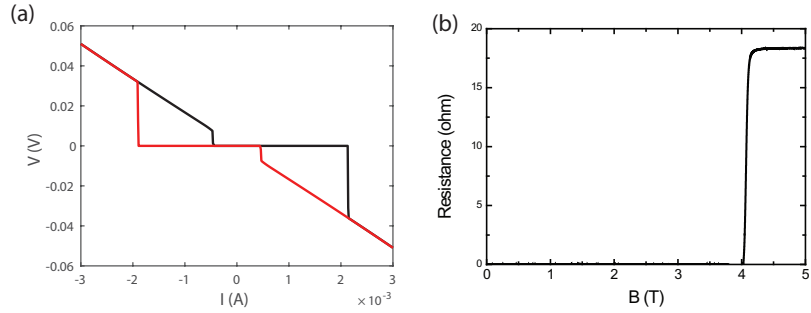

**Supplementary Figure 1: Characterization of superconductivity in Niobium.** As mentioned in the manuscript a Nb channel was fabricated using electron beam lithography followed by 30 nm deposition of Niobium by DC sputtering. The channel length was 5 $\mu$m and width was 2 $\mu$m. (**a**) The I versus V characteristic of the Nb channel at 240 mK showing the supercurrent. Black is the forward (negative current to positive current) scan and red is the reverse scan. (**b**) The resistance is plotted as a function of magnetic field at 240 mK, where one can clearly see the critical magnetic field, $B_C \sim 4$ T. The resistance was also measured similarly as a function of temperature and the critical temperature was found to be, $T_C \sim 7$ K.

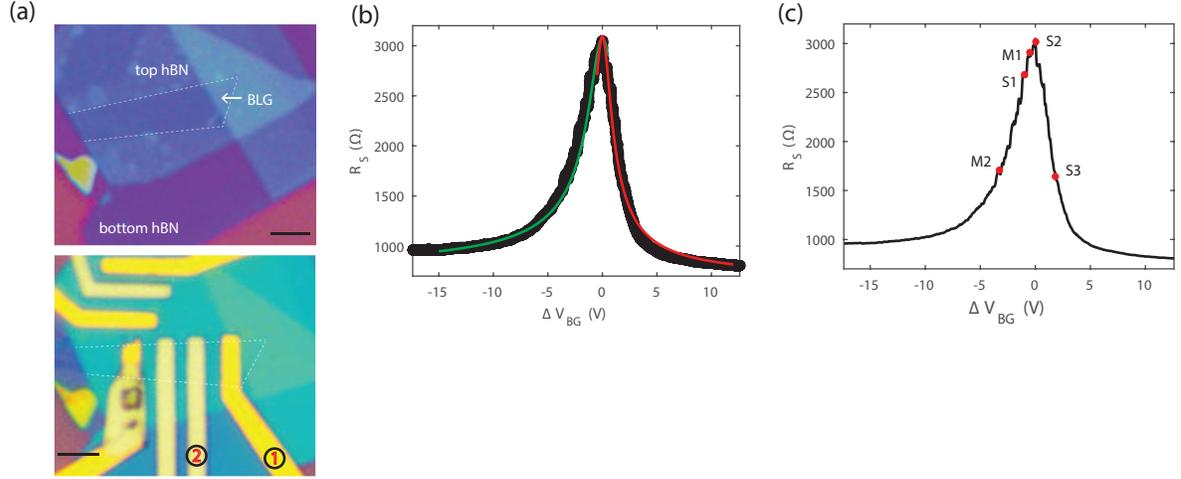

**Supplementary Figure 2: Fabrication and characterization of BLG - Nb device.** The edge contacted BLG device was fabricated using standard dry transfer technique as discussed in the manuscript. (**a**) The encapsulated BLG device before (top) and after (bottom) making contacts with scale bar of 2 $\mu$m. All the measurements were performed between contact-1 (Au) and contact 2 (Nb). (**b**) Gate response of the device (black) at 40 mK using standard lock-in technique. The experimental data is fitted with the equation $R_S = R_C + \frac{L}{We\mu\sqrt{(C\Delta V_{BG})^2 + n_0^2}}$, where $C$ and $n_0$ are capacitance per unit area and carrier inhomogeneity respectively. Fitting in the electron side (red) gives a mobility ($\mu$) $\sim$ 15000 $cm^2/V.sec$ and total contact resistance ($R_C$) $\sim$ 700 $\Omega$. Similarly, fitting in the hole side (green) gives $\mu \sim$ 10000 $cm^2/V.sec$ and $R_C \sim$ 800 $\Omega$. (**c**) M1 ($\Delta V_{BG}$ = -0.4 V) and M2 (-3.4 V) are the locations where the shot noise data are shown in the manuscript, whereas S1 (-1.2 V), S2 (0 V) and S3 (2 V) are the locations where the shot noise data are shown in the supplementary information.

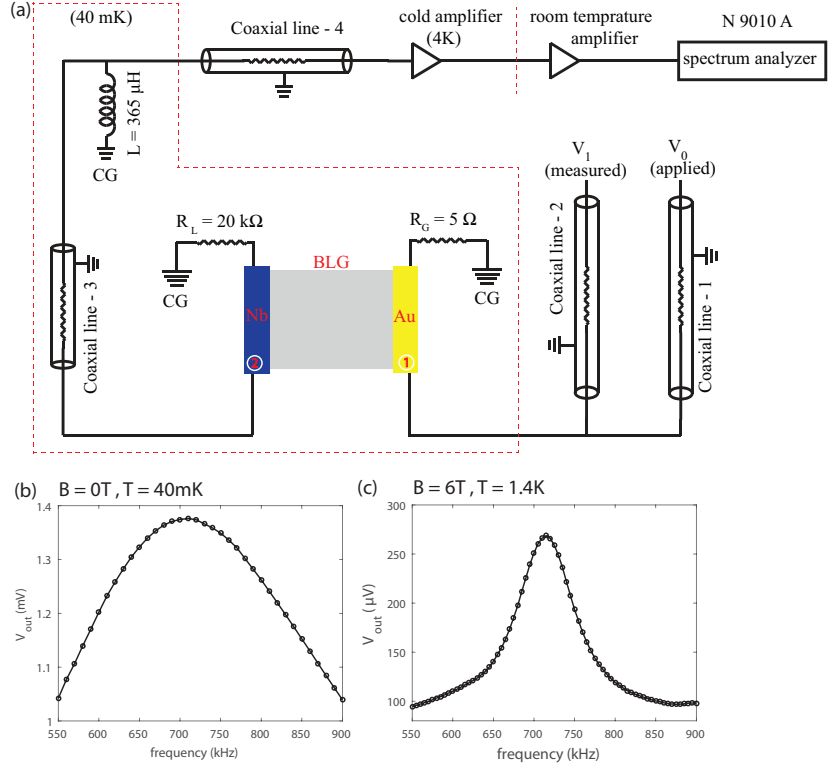

**Supplementary Figure 3: Measurement setup.** (**a**) The schematic of the detailed shot noise measurement setup. At the input end (contact-1) the sample is biased by a voltage across the 5Ω resistance, which was grounded at 40 mK plate of the fridge. The grounds at 40 mK plate are called cold ground (CG). The coaxial line - 1 is used to apply the voltage $V_0$, whereas another coaxial line (line-2) is used to measure the exact voltage across the 5 Ω resistor $V_1$. The contact-2 is connected to cold ground through a $R_L = 20$ kΩ resistor parallel to an superconducting inductor with inductance of L ∼ 365 $\mu$H, which is then connected to the cold amplifier placed at 4 K plate of the dilution fridge. The coaxial lines (line-3 and line-4) add up to a total capacitance of C ∼ 138 pF. The inductor (L) and capacitor (C) makes a resonant circuit with a resonance frequency, $f_0$ ∼ 710 kHz. The bandwidth of the resonant circuit depends on the parallel effective resistance ($R_{eff}$) of sample resistance ($R_S$) and load resistance ($R_L$). The voltage fluctuation at contact-2 was amplified by the cold amplifier and a room temperature amplifier, and finally measured by a spectrum analyzer at the resonance frequency with a bandwidth (Δf) of 30 kHz. The resistance of coaxial lines (line-1 and line-2) was $R_{Line}$ ∼ 65 Ω, whereas the capacitance of each coaxial line was $R_{Line}$ ∼ 425 pF. Coaxial line - 3 and coaxial line -4 are of quite short length with negligible resistance. (**b**) Output signal measured at the spectrum analyzer ($V_{OUT}$) is plotted as a function of frequency with $V_1 = 1.26$ $\mu$V at the Dirac point ($\Delta V_{BG} = 0$ V) showing the resonance behavior with $f_0$ ∼ 710 kHz for B=0 T. (**c**) The total noise coming from thermal as well as cold amplifier measured at the spectrum analyzer ($V_{OUT}$) is plotted as a function of frequency at the Dirac point for B=6 T and T=1.4 K, which also shows the expected resonance behavior with $f_0$ ∼ 710 kHz.

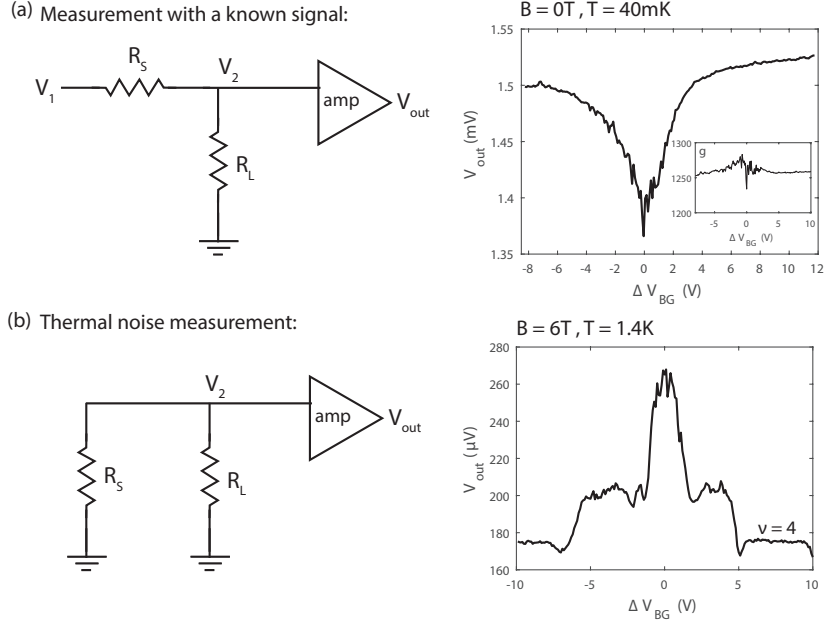

**Supplementary Figure 4: Gain calibration.** The most challenging part of shot noise measurement technique employing RLC resonance circuit and cold amplifier is extracting the accurate gain of the cold amplifier, which can change in each cool down of the fridge. One of the ways is, by applying a known signal at resonance frequency at contat-1 ($V_1$) and by knowing the sample resistance ($R_S$) from Supplementary Fig. 2b, the output signal at the spectrum analyzer ($V_{OUT}$) can be used to extract the total gain (g) as $V_{OUT} = \frac{V_1 R_L g}{R_S + R_L}$. Effective circuit for the measurement with known signal at resonance frequency is shown in **a**-left panel. $V_{OUT}$ plotted as a function of $\Delta V_{BG}$ measured at 710 kHz with $V_1$ = 1.26 $\mu$V at T=40 mK and B=0 T (**a**-right panel). The inset shows the extracted gain, g $\sim$ 1260, which varies within 5% as a function of $\Delta V_{BG}$. Another way of measuring total gain without applying a known signal is by thermal noise measurement. The measured total noise at a given temperature T is given by $V_{out}^2(T) = (4k_B T R_{eff} + S_i^{amp} R_{eff}^2 + S_v^{amp}) \times \Delta f \times g^2$. By measuring the total noise at two different temperatures ($T_1$ and $T_2$) the total gain can be extracted as $g^2 = \frac{V_{out}^2(T_2) - V_{out}^2(T_1)}{4k_B R_{eff}(T_2 - T_1)\Delta f}$. Effective circuit for the thermal measurement at the resonance frequency is shown in **b**-left panel. $V_{OUT}$ plotted as a function of $\Delta V_{BG}$ measured at 710 kHz at T=1.4 K and B=6 T (**b**-right panel). Similarly the $V_{OUT}$ was measured at T=40 mK. Using the values of $V_{OUT}$ in the $\nu$ = 4 QH plateau at 40 mK and 1.4 K, we extracted the total gain, g $\sim$ 1330. Both of the measurements gives very similar values of g. We have used g = 1300 for all the analysis of shot noise data.

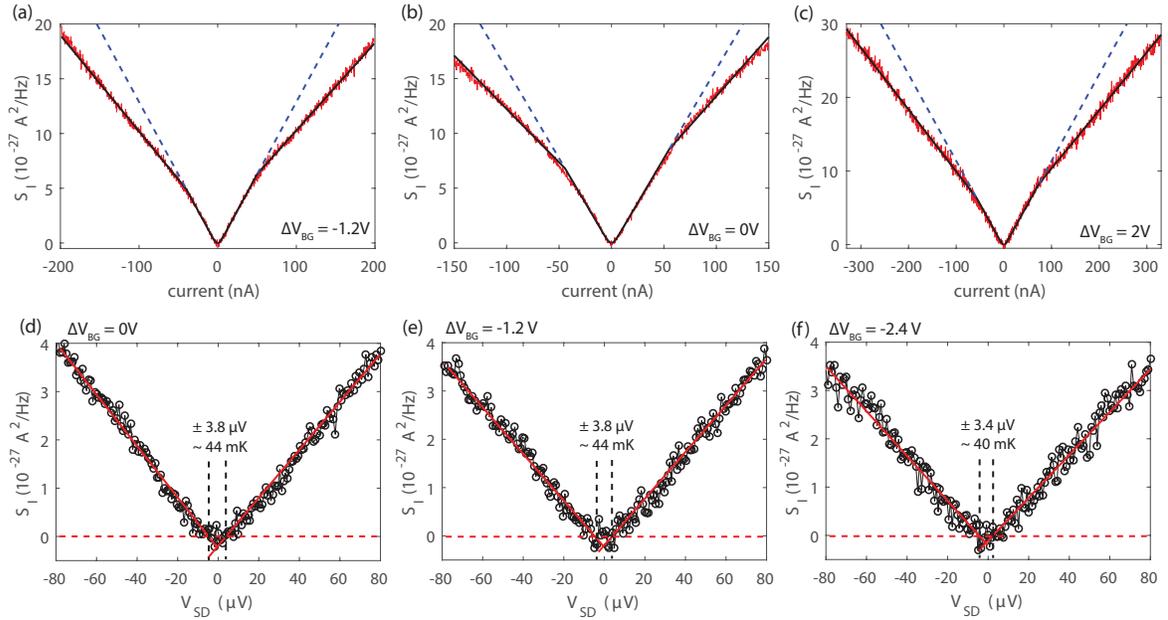

**Supplementary Figure 5: Additional shot noise data at zero magnetic field.** (**a - c**) Additional shot noise data at zero magnetic field plotted as a function of applied DC current at gatevoltages marked by S1, S2 and S3 in Supplementary Fig. 2c. Two distinct slopes are clearly observed in all the data signifying transport by AR at lower excitation energy. (**d - f**) Electron temperature is extracted from the shot noise data at several gatevoltages using the relation e*$V_{ext}$ = 2$k_B$$T_e$, where e*=2e and $V_{ext}$ is the voltage at which the extrapolation of linear part of the shot noise data intersects the $S_I$=0 line. Though the shot noise data in these figures have lot of fluctuations, the fitting of shot noise using equation 2 of manuscript with T as a fitting parameter also gives similar temperature, around 35-45 mK.

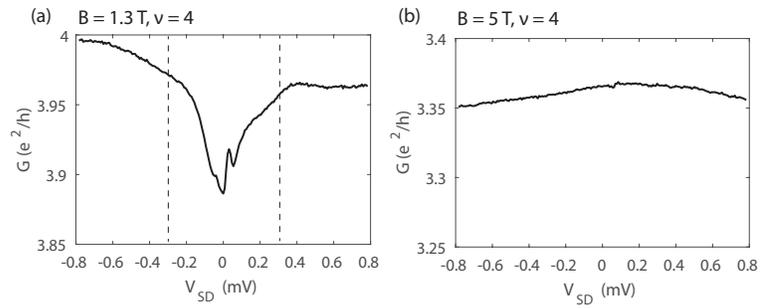

**Supplementary Figure 6: BCS like feature in the bias response in QH regime.** (**a-b**) Conductance plotted as a function $V_{SD}$ for the $\nu$=4 plateau at B=1.3 T in (a) showing the BCS like features which is not present at B=5T (>$B_C$) in (b). Note that the conductance data at 5T contains the Nb lead resistance of $\sim 500\ \Omega$.

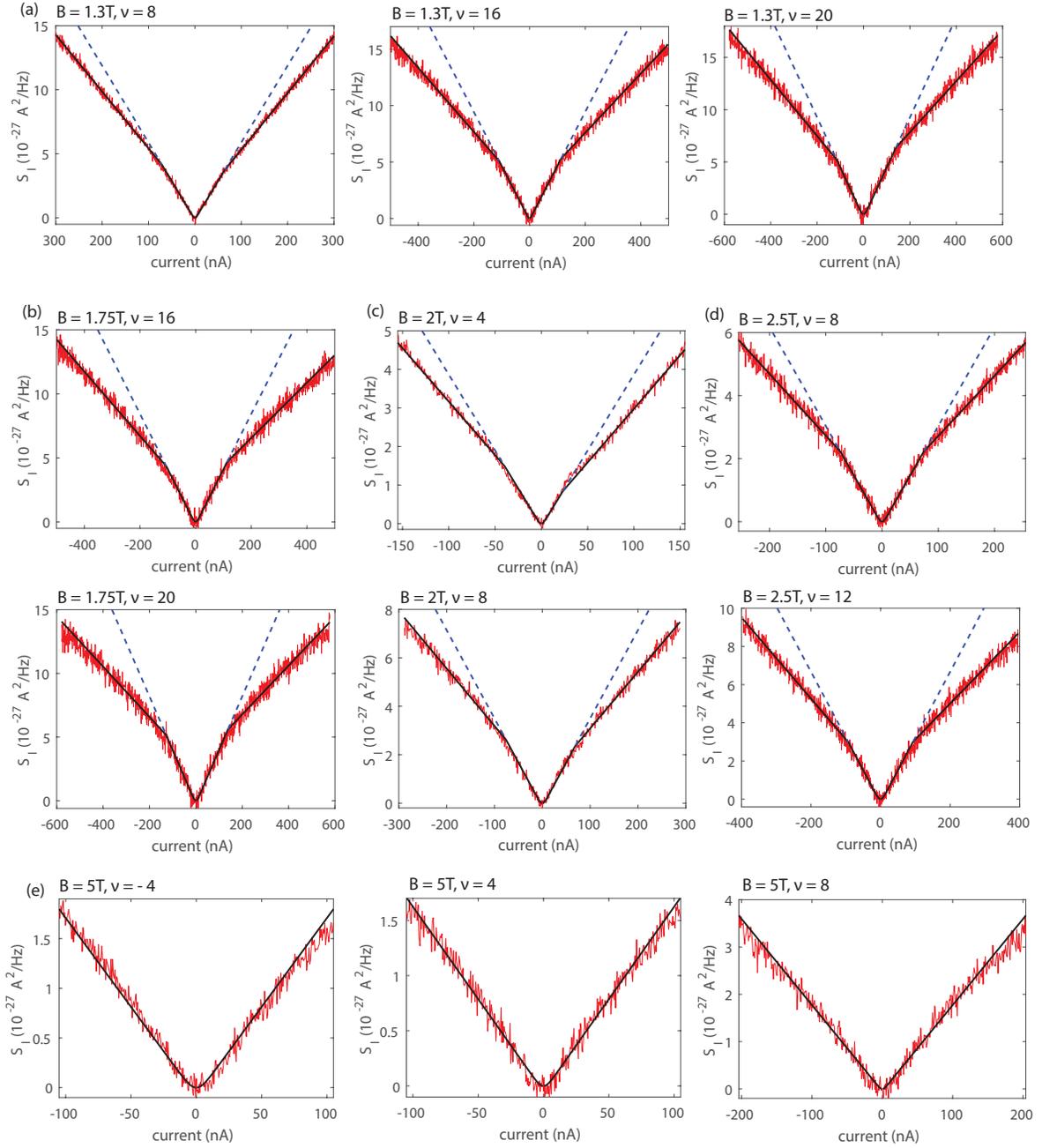

**Supplementary Figure 7: Additional shot noise data in QH regime.** (**a - e**) Shot noise data at several magnetic fields and filling factors showing two distinct slopes below critical magnetic field. At 5T there is single slope as expected for normal transport.


\end{document}